\documentclass[twocolumn,amsmath,amssymb,nofootinbib]{revtex4}
\usepackage{graphicx}% Include figure files
\usepackage{dcolumn}% Align table columns on decimal point
\usepackage{bm}% bold math
\usepackage{parskip}
\usepackage{color}
\usepackage{subfigure}
\usepackage{amssymb}
\usepackage{slashed}
\usepackage[export]{adjustbox}

\usepackage{amsmath}
\usepackage{amssymb}
\usepackage{graphicx}

\def\be{\begin{eqnarray}}
\def\ee{\end{eqnarray}}
\def\bea{\begin{eqnarray}}
\def\eea{\end{eqnarray}}

\begin{document}

\title{A Combined Approach to the Analysis of Space and Ground Experimental Data \\[0.25cm]
Within a Simplified E$_6$SSM}

\author{Shaaban Khalil$^{1}$, Kamila Kowalska$^2$, Stefano Moretti$^3$, Diana Rojas-Ciofalo$^2$, and  Harri Waltari$^{3,4}$}
\vspace*{0.2cm}
\affiliation{$^1$ Center for Fundamental Physics, Zewail City of Science and Technology, 6 October City, Giza 12588, Egypt\\
$^2$ National Centre for Nuclear Research, Pasteura 7, 02-093 Warsaw, Poland\\
$^3$ School of Physics and Astronomy, University of Southampton,
	Southampton, SO17 1BJ, UK\\
$^4$ Particle Physics Department, Rutherford Appleton Laboratory, Chilton, Didcot, Oxon OX11 0QX, UK}

%\email[]{skhalil@zewailcity.edu.eg}
%\email[]{kamila.kowalska@ncbj.gov.pl}
%\email[]{s.moretti@soton.ac.uk}
%\email[]{dianitzdr@gmail.com}
%\email[]{h.waltari@soton.ac.uk}

\date{\today}

\begin{abstract}
Within the Exceptional Supersymmetric Standard Model (E$_6$SSM),  we investigate striking signatures at the Large Hadron Collider (LHC) for a  long-lived charged inert higgsino, which is degenerate with the inert neutralino at tree level and a mass splitting of order ${\cal O}(0.3)$ GeV is generated at the loop level, resulting in a lifetime of order ${\cal O}(0.02)$ nanoseconds.
We focus on the most sensitive search for long-lived charged inert higgsino decays to the lightest neutral inert higgsino Dark Matter (DM) and very soft charged leptons, which are eventually stopped in the detector resulting in a disappearing-track signal.
Furthermore, we study the displaced vertex signature of the inert chargino in the case where it is produced via the $Z^{\prime}$ portal. We illustrate how difficult it is to construct displaced vertices in this class of models, though some evidence of these could be gained at the High Luminosity LHC (HL-LHC). Finally, we compare the spin independent and spin dependent cross sections of the lightest inert higgsino DM to those of current direct detection experiments, proving that it is possible to gain sensitivity to the active DM component of this scenario in the near future. The combination of these signatures with the  one emerging from $Z'$ production and decay via Drell-Yan (DY), which can be characterised as belonging to the E$_6$SSM via both the cross section and Forward-Backward Asymmetry ($A_{\rm FB}$), could point uniquely to this non-minimal realisation of Supersymmetry (SUSY).  In fact, we remark that these evidences of the E$_6$SSM can also be correlated to the features of standard SUSY cascades in terms of the amount of Missing Transverse Energy (MET or $\slashed{E}_T$) present therein.

\end{abstract}

\maketitle
%

%%%%%%%%%%%%%%%%%%%%%%%%%%%%%%%%%%%%%%%%%%%%%%%%%%%%%%%%%%%%%%%%%%%%%%%%%%%%%%%%%%
\section{Introduction}

The E$_6$SSM is a Supersymmetry (SUSY) extension of the Standard Model (SM)  inspired by string theory, with an exceptional gauge unification group of  E$_6$ type \cite{King:2005jy}--\cite{King:2020ldn}. This model provides a natural framework to account for neutrino masses and  solving the $\mu$-problem of SUSY. The high scale E$_6$ symmetry can be spontaneously broken down to $SU(3)_C \times SU(2)_L \times U(1)_Y \times U(1)'$, where $ U(1)' = \cos \theta~ U(1)_\chi + \sin \theta~ U(1)_\psi$, with $U(1)_\chi$ and $U(1)_\psi$ are two anomaly free $U(1)$'s and $\tan \theta = \sqrt{15}$. Right-handed neutrinos are identified as singlet components of the fundamental representation of E$_6$ $27_i$-plets, $i=1,2,3$, which are also not charged under the extra $U(1)'$. As a result, they may acquire heavy Majorana masses, necessitating a large scale seesaw mechanism (see~\cite{Moretti:2019ulc} for a detailed treatment of the necessary dynamics and various model realisations within SUSY).

Recently, a simplified E$_6$ Supersymmetric Standard Model (E$_6$SSM) has been studied~\cite{Khalil:2020syr}, with a focus on possible multi-component Dark Matter (DM). 
This type of simplified E$_6$SSM is governed by a number of discrete symmetries (for definitions, see \textit{e.g.} \cite{Hall:2011zq}). In particular, $Z_2^H$ that distinguishes the third generation of active Higgs doublets from other (inert~\footnote{We acknowledge here the somewhat misleading use of the word `inert', especially when referring to electrically charged objects. However, we have decided to adhere to such a nomenclature (see later for the corresponding definition) as it is well established in the literature, e.g., see \cite{Moretti:2019ulc} and references therein.}) Higgs doublets, $Z_2^L$ or $Z_2^B$  to prevent fast proton decay and $Z_2^M\equiv R$-parity to avoid the $B-L$ violating terms in the Superpotential.  These symmetries are critical in suppressing Flavour Changing Neutral Currents (FCNCs). It has been demonstrated that stable active and inert higgsino particles that can be candidates for DM are quite natural in this class of models. In addition, future $e^+e^-$ collider probes of these particles have been also studied \cite{Khalil:2021ycm}.  Finally, their potential signatures in direct and indirect detection experiments have been investigated \cite{Khalil:2020syr}.
 
In this article, we look at other promising signals for this type of a model, based on the Long Lived (LL) inert charged higgsino ($ \tilde{\chi}_I^\pm$), which is quite degenerate with the neutral inert higgsino ($ \tilde{\chi}_I^0$). The lifetime of the inert charged higgsino can be of the order of ${\cal O}(0.02)$ nanoseconds and upwards, indicating that it is a LL Particle (LLP), with striking signatures at the Large Hadron Collider (LHC).  We will focus on the disappearing-track signature, which offers the most sensitive search for a LLP charged inert higgsino in the case where it decays to the lightest neutral inert higgsino and soft charged particles. In addition, we will look into the a displaced vertex signature arising from the inert chargino produced through the $Z^{\prime}$ portal. We emphasise that the construction of these displaced tracks is very challenging. Naturally, the $Z^{\prime}$ may also be discovered through other decay channels and we discuss this possibility too, specifically, in Drell-Yan (DY) channels. We also illustrate the spin independent and spin dependent cross sections of the active and inert higgsino DM and compare the results to those of current direct detection experiments.

Before introducing the reader to the layout of the paper, we would like to stress that the novelty of our work resides primarily  in the
{\sl simultaneous} access to the variety of SUSY discovery channels that we will discuss here (some of which specific to the E$_6$SSM, while others in common with simpler realisations). This multi-prong approach, attempted here not only across a variety of signatures but also at different (space and ground)  experiments, has never been adopted  before. Indeed, we deem that, the more complicated the SUSY realisation considered, the more this is necessary.

Following this introduction, 
the plan for the remainder of the paper is as follows. In the next section, we recap the salient features of the E$_6$SSM. Then we discussed the aforementioned experimental signatures of it. Finally, we conclude.

%%%%%%%%%%%%%%%%%%%%%%%%%%%%%%%%%
\section{The model}\label{sec:model}
As mentioned, the fundamental representation of E$_6$ is a $27_i$-plet, $i=1,2,3$, which has the following decomposition under $SU(5) \times U(1)'$: 
\bea 
27_i &\to& (10, \frac{1}{\sqrt{40}})_i + (\bar{5}, \frac{2}{\sqrt{40}})_i + (\bar{5}, \frac{-3}{\sqrt{40}})_i  + (5, \frac{-2}{\sqrt{40}})_i \nonumber\\
&+& (1, \frac{5}{\sqrt{40}})_i + (1,0)_i ,
\eea
where the following field associations can be made: $(10, \frac{1}{\sqrt{40}})_i$ and $(\bar{5}, \frac{2}{\sqrt{40}})_i$ are the normal matter, $(\bar{5}, \frac{-3}{\sqrt{40}})_i$ and $(5, \frac{-2}{\sqrt{40}})_i$ are three generations of Higgs doublets $H_{di}, H_{ui}$ and exotic coloured states $\bar{D}_i, D_i$, $(1, \frac{5}{\sqrt{40}})_i$ are three generations of singlets $S_i$ and $(1,0)_i$ are the right-handed neutrinos. We assume that the exotic matter ($\bar{D}_i, D_i$) is heavy and consider the effective theory where these particles have been integrated out. The discrete symmetries mentioned above force the first two generations of Higgs doublets and scalar singlets to be inert, with vanishing Vacuum Expectation Values (VEVs). In this regard, the $U(1)'$ is spontaneously broken by the singlet, $S_3$, which radiatively develops a VEV: $\langle S_3\rangle = \frac{s}{\sqrt{2}}$, resulting in a mass for the $Z'$ gauge boson of the order of the SUSY breaking scale, say, a few TeVs. While the VEVs of third generation Higgs doubles, $\langle H^0_{d3}\rangle = \frac{v_d}{\sqrt{2}} = \frac{v\cos\beta}{\sqrt{2}}$ and $ \langle H^0_{u3}\rangle = \frac{v_u}{\sqrt{2}} = \frac{v\sin\beta}{\sqrt{2}}$, spontaneously break the electroweak symmetry.

The superpotential of our simplified E$_6$SSM,  with the above mentioned symmetries, is given by 
\bea
W &=& Y_{u} Q U^c H_u + Y_{d} Q D^c H_d + Y_{e} L E^c H_d + Y_{\nu} L \nu^c H_u  \nonumber\\
&+& \lambda S H_d H_u.
\label{Superpot}
\eea
where $\lambda S H_d H_u$ stands for $\lambda_{ijk} S_i H_{d_j} H_{u_k}$. Thus, the $\mu$ term is generated dynamically by  the VEV of the singlet $S_3$ and is given by $\mu_{\mathrm{eff}} = \lambda_{333} \frac{s}{\sqrt{2}}$.  Since $s$ is of order the SUSY breaking scale, the $\mu$ term is of the desired  TeV scale. 

The salient feature of this class of models is the existence of inert spectrum (inert Higgs bosons, inert higgsinos, and inert charginos) in addition to the usual MSSM spectrum.  
On the basis of \( \left(\tilde{\chi}^{-,I}_{1}, \tilde{\chi}^{-,I}_{2}\right)\), where $\tilde{\chi}^{-,I}_{1(2)}=\left(\tilde{h}^{-,I}_{d(u)_{1}}, \tilde{h}^{-,I}_{d(u)_{2}}\right)$, one can find the mass matrix for inert charginos as follows:
\begin{equation} 
M_{\tilde{\chi}^{\pm}_I} = \left( 
\begin{array}{cc}
-\frac{1}{\sqrt{2}}v_s\lambda_{311} & -\frac{1}{\sqrt{2}}v_s\lambda_{312} \\ 
-\frac{1}{\sqrt{2}}v_s\lambda_{321} & -\frac{1}{\sqrt{2}}v_s\lambda_{322} 
\end{array} 
\right). 
\end{equation} 
Also, the mass matrix for the inert higgsinos (neutralinos) in the basis of \( \left(\tilde{h}^{0,I}_{d1}, \tilde{h}^{0,I}_{d2}, \tilde{h}^{0,I}_{u1}, \tilde{h}^{0,I}_{u2}\right)\) is given by 
{\small \begin{equation} 
M_{\tilde{\chi}^{0}_I} = \left( 
\begin{array}{cccc}
0 & 0 & -\frac{1}{\sqrt{2}}v_s\lambda_{311} & -\frac{1}{\sqrt{2}}v_s\lambda_{312} \\ 
0 & 0 & -\frac{1}{\sqrt{2}}v_s\lambda_{321} & -\frac{1}{\sqrt{2}}v_s\lambda_{322} \\ 
-\frac{1}{\sqrt{2}}v_s\lambda_{311} & -\frac{1}{\sqrt{2}}v_s\lambda_{312} & 0 & 0 \\ 
-\frac{1}{\sqrt{2}}v_s\lambda_{321} & -\frac{1}{\sqrt{2}}v_s\lambda_{322} & 0 & 0 
\end{array} 
\right). 
\end{equation}}
At tree-level the inert charginos and neutralinos are  degenerate with a mass $\lambda_{3ij}s/\sqrt{2}$. A mass splitting $m_{\tilde{\chi}_I^\pm}-m_{\tilde{\chi}_I^0}< 1$~GeV is however generated through loop corrections, where a lower case  denotes the mass of the lightest inert chargino and neutralino(s), respectively.

It was emphasised in  \cite{Khalil:2020syr}  that the striking signature of $E_6$SSM is the possibility of having two-component DM. One of these DM components was found to be the lightest active neutralino (higgsino-like), $\tilde{\chi}_1$,  with direct couplings to the SM fermions. The other DM component is the lightest inert neutralino (inert-higgsino-like) that does not interact directly with the SM fermions. These two particles are stable and have the potential to play an important role in accounting for the DM in the Universe.

In \cite{Khalil:2020syr} some of us noted that for a large part of the available mass range the sum of the two higgsino masses is nearly constant, when the relic density gets its observed value. Hence a light active higgsino means a heavy ($\sim$TeV) inert higgsino and vice versa. We will be here mostly interested in the case, where the inert higgsino is light so that it could be detectable at colliders and hence the active higgsino is in the TeV-range, outside the reach of current colliders but could potentially be probed in direct detection experiments.

As we are interested here in analyzing the signatures of innert charged higgsino as LLP at the LHC, we will provide the relevant interactions of this particle involved in its production and decay. In our calculations, we use the following interaction terms:
\be 
{\cal L} \!=\! \tilde{\chi}^{+}_I Y_1 \tilde{\chi}^{0}_I W^- \!+\!   \tilde{\chi}^{+}_I Y_2 \tilde{\chi}^{-}_I Z \!+\!  \tilde{\chi}^{+}_I Y_3 \tilde{\chi}^{-}_I Z' \!+\! h.c., 
\ee
where $Y_{1,2,3}$ are given by 
\begin{widetext}
\bea 
Y_1 &=& -i \frac{1}{\sqrt{2}} g_2 \gamma_\mu P_L \sum_{a=1}^2  \left(Z_{ja}^{\tilde{h}^-}\right)^{*} Z_{ia}^{\tilde{h}^0} 
+  i \frac{1}{\sqrt{2}} g_2 \gamma_\mu P_R  \sum_{a=1}^2  \left(Z_{i2+a}^{\tilde{h}^0}\right)^{*} Z_{i2+a}^{\tilde{h}^+}.  \\
Y_2 &=& \frac{i}{20} \delta_{ij} \Big(-10 g_1 \cos \theta'_W \sin \theta_W + 10 g_2   \cos \theta'_W \cos \theta_W  + 2 \sqrt{10} g_N \sin \theta'_W \Big) \gamma_\mu P_L \nonumber\\
&+& \frac{i}{10} \delta_{ij} \Big(-5 g_1 \cos \theta'_W \sin \theta_W +  5 g_2   \cos \theta'_W \cos \theta_W  -  \sqrt{10} g_N \sin \theta'_W \Big) \gamma_\mu P_R.\\
Y_3 &=&  \frac{i}{20} \delta_{ij} \Big(10 \left( g_1 \sin \theta'_W  -g_2 \cos\theta_W \right) \sin \theta'_W + 3 \sqrt{10}  g_N   \cos \theta'_W \Big) \gamma_\mu P_L \nonumber\\
&-&   \frac{i}{20} \delta_{ij}  \Big(5 \left( -g_1 \sin \theta_W  + g_2 \cos\theta_W \right) \sin\theta'_W + \sqrt{10} g_N \cos\theta'_W\Big) \gamma_\mu P_R.\\
\eea
\end{widetext}
Here $Z^{\tilde{h}^-}$ and $Z^{\tilde{h}^0}$ stand for the diagonalising matrices of inert charged and neutral higgsinos, respectively, while $g_1$, $g_2$ and $g_N$  are the $U(1)_Y$, $SU(2)_L$ and $U(1)'$ gauge couplings. 

%%%%%%%%%%%%%%%%%%%%%%%%%%%%%%%%%%%%%%

\section{Experimental signatures}

There are several possible signatures that can arise at the LHC or at other contemporary experiments. The most obvious one is the existence of a new gauge boson $Z^{\prime}$, which presumably would first be seen as a dilepton or dijet resonance. The possibility of seeing this resonance depends mostly on the  $Z^{\prime}$ mass. Besides the mass, the second important parameter is the kinetic mixing between the two $U(1)$ gauge bosons, which can lead to a substantial Branching Ratio (BR) for the decay $Z^{\prime}\rightarrow W^{+}W^{-}$. As such a BR can be up to $90\%$ or so the $Z^{\prime}$ can become a wide resonance and  the BRs of all other decay modes are suppressed, including those into  Superpartners, which are generally subleading. In such circumstances the dilepton and dijet signatures might not be visible against the SM background. 

This simplified E$_6$SSM has two DM candidates, one from the active sector and one from the inert sector of the model. Direct detection experiments are mostly sensitive to the active component. The neutralino-neutralino-$Z$ coupling depends on the mixing of the neutralino sector. Current experiments can already rule out data points, where the mixing between the gauginos and higgsinos is large. This happens when $||M_{1}|-|\mu_{\mathrm{eff}}||$ or $||M_{2}|-|\mu_{\mathrm{eff}}||$ is less than $\mathcal{O}(200)$~GeV. The inert sector DM candidate has a cross section below the neutrino floor. Neither of the components give rise to indirect detection signatures that would be detectable \cite{Khalil:2020syr}.

Hence the only way to detect the inert sector will be collider experiments. We have already pointed out the chance of mono-photon signatures at electron-positron colliders \cite{Khalil:2021ycm}. At the LHC the best possibility of searching nearly degenerate charginos and neutralinos are disappearing tracks. As the majority of (neutralino-)chargino pair production occurs via SM gauge bosons, the chances of seeing disappearing tracks depends almost solely on the chargino/neutralino mass.

There is also a chance of producing inert chargino pairs via the $Z^{\prime}$ portal. If the $Z^{\prime}$ is heavy enough compared to the inert chargino, the charged leptons that arise from the chargino decay could be boosted so much that they survive to the calorimeters and muon detectors. In such a case,  we would have tracker hits  implying large momentum while seeing a soft lepton in the same direction.

Naturally standard searches for Superpartners are also sensitive to the squark, slepton and gaugino sectors of this model. Such signatures would be
Minimal Supersymmetric Standard Model  (MSSM)-like and hence distinguishing our model from the latter would require the discovery of the $Z^{\prime}$ or the inert sector.

\begin{table}[b]
\begin{tabular}{l| c c c}
\hline
\hline
 & BP1 & BP2 & BP3\\
\hline
$m(Z^{\prime})$ & 4212 & 4255 & 5359\\
$\Gamma(Z^{\prime})$ & 86 & 88 & 118\\
BR$(Z^{\prime}\rightarrow W^{+}W^{-})$ & $2\%$ & $2\%$ & $5\%$\\
$m(\tilde{\chi}_{I}^\pm)$ & 242 & 273 & 290\\
$\Delta m_{I}$ & 0.34 & 0.35 & 0.35\\
$m(\tilde{\chi}_{1})$ & 1155 & 1135 & 1089\\
\hline
\hline
\end{tabular}
\caption{The benchmark points used in our study. We give the masses of the $Z^{\prime}$, the lightest inert chargino and the lightest active neutralino in GeV's. We choose $g_{N}=0.55$ for all benchmarks. The mass splitting $\Delta m_{I}$ is that between the inert chargino and neutralino. All benchmarks satisfy the relic density constraint $\Omega h^{2}=0.120\pm 0.002$.\label{tb:benchmarks}}
\end{table}

\begin{figure*}[!t]
\hspace{0cm}\includegraphics[scale=0.3,angle=90]{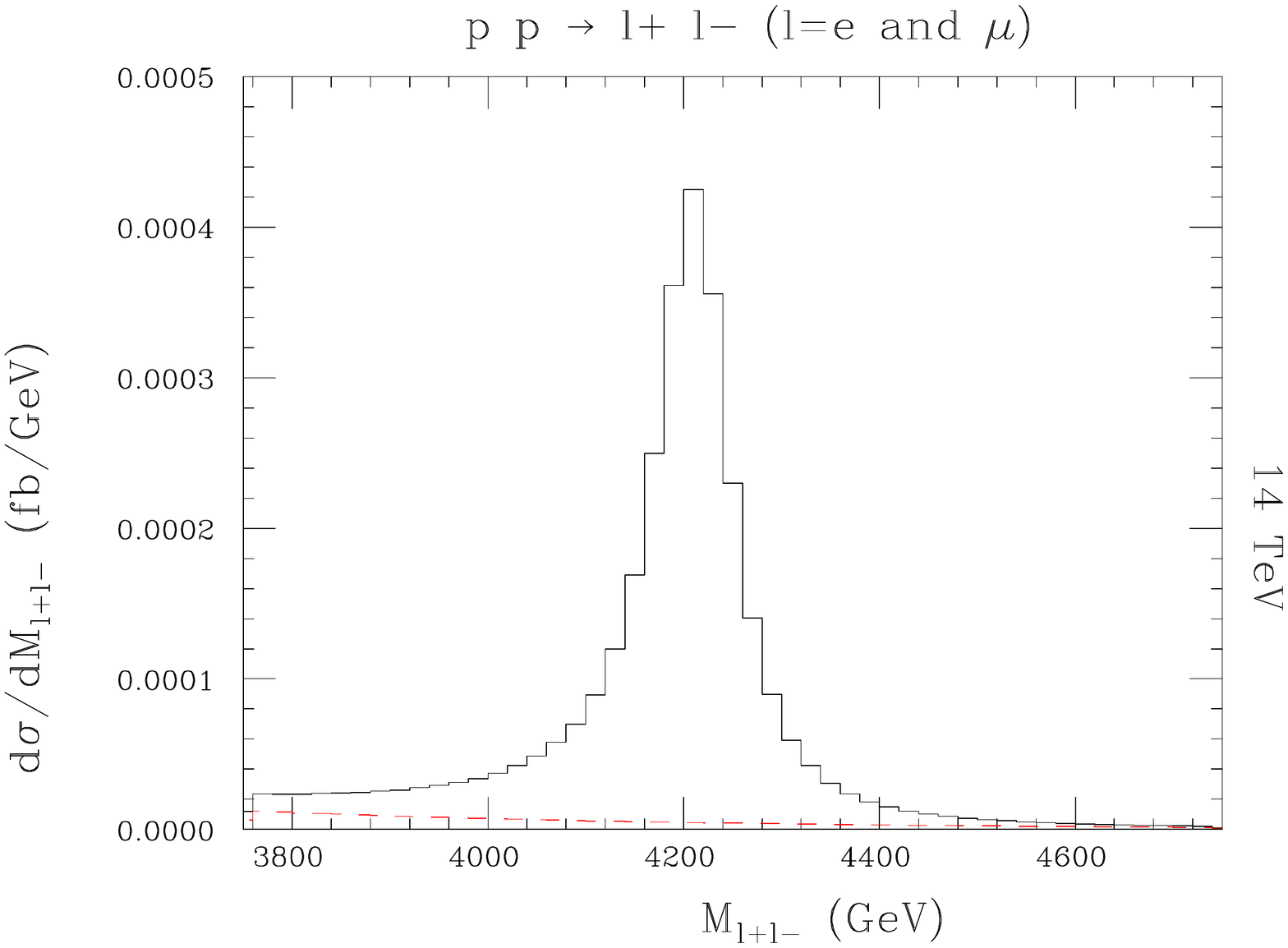} %\hspace*{1.truecm}
\hspace{0cm}\includegraphics[scale=0.3,angle=90]{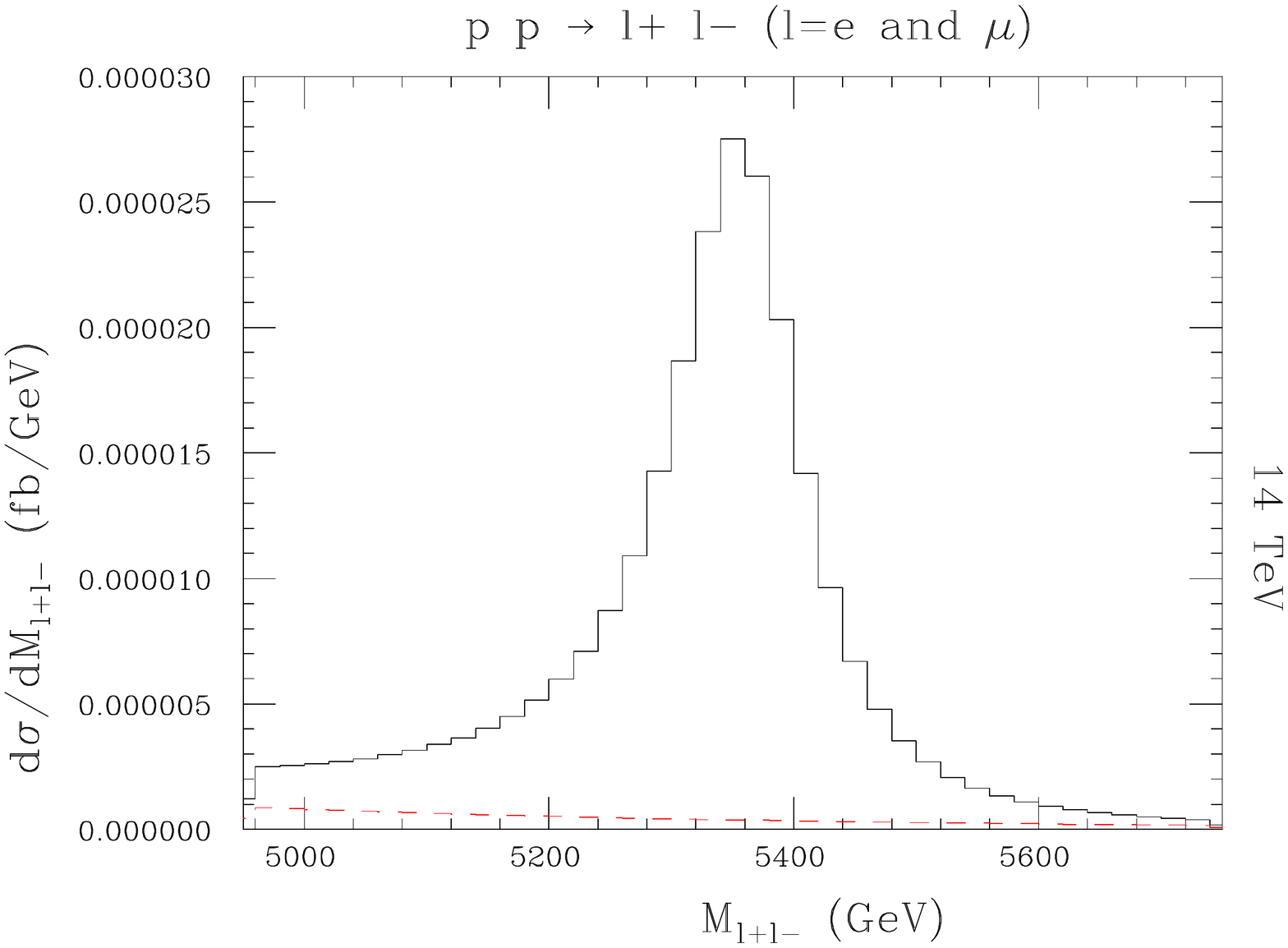} \\
\hspace{0cm}\includegraphics[scale=0.3,angle=90]{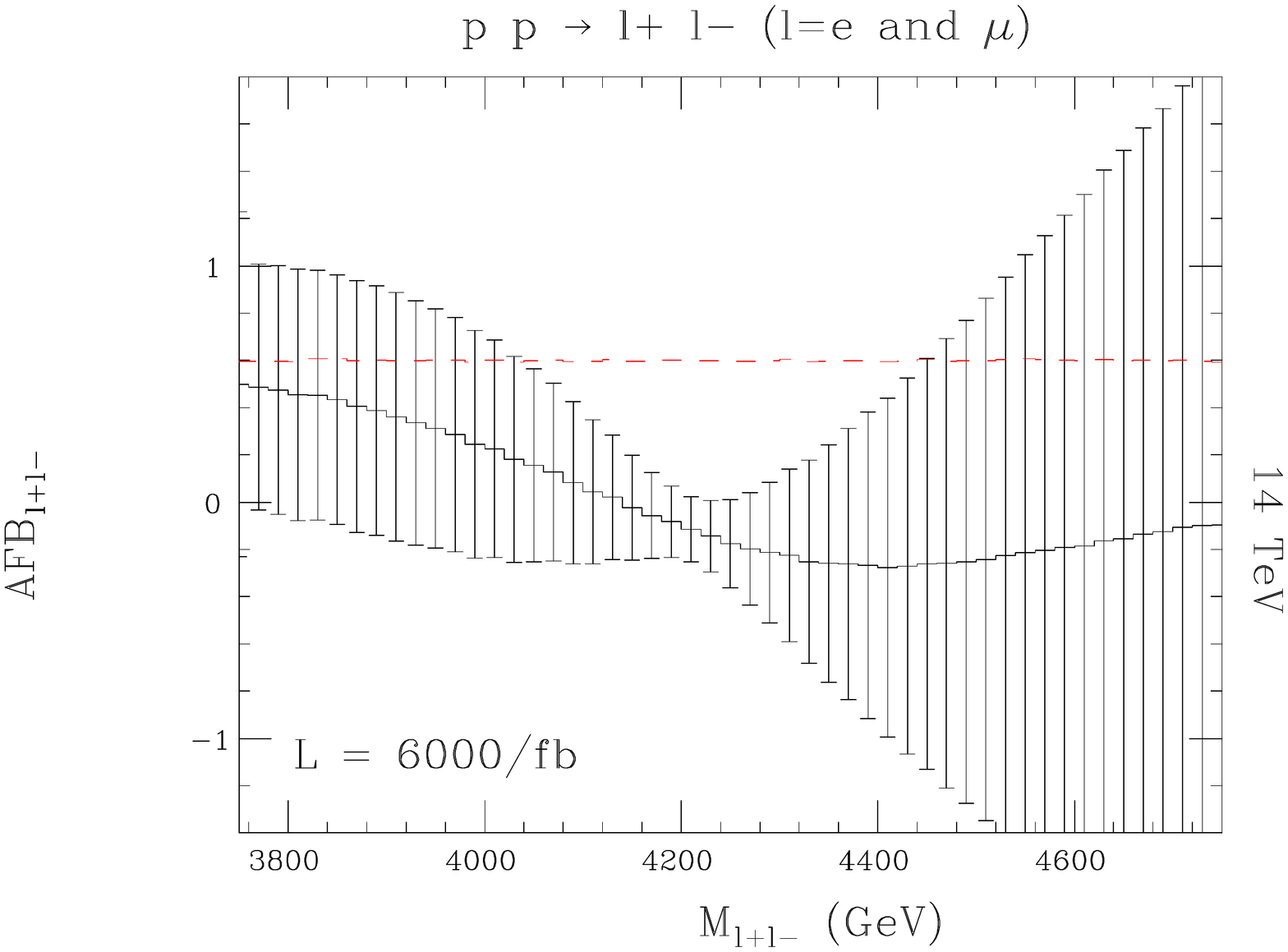} %\hspace*{3.truecm}
\hspace{0cm}\includegraphics[scale=0.3,angle=90]{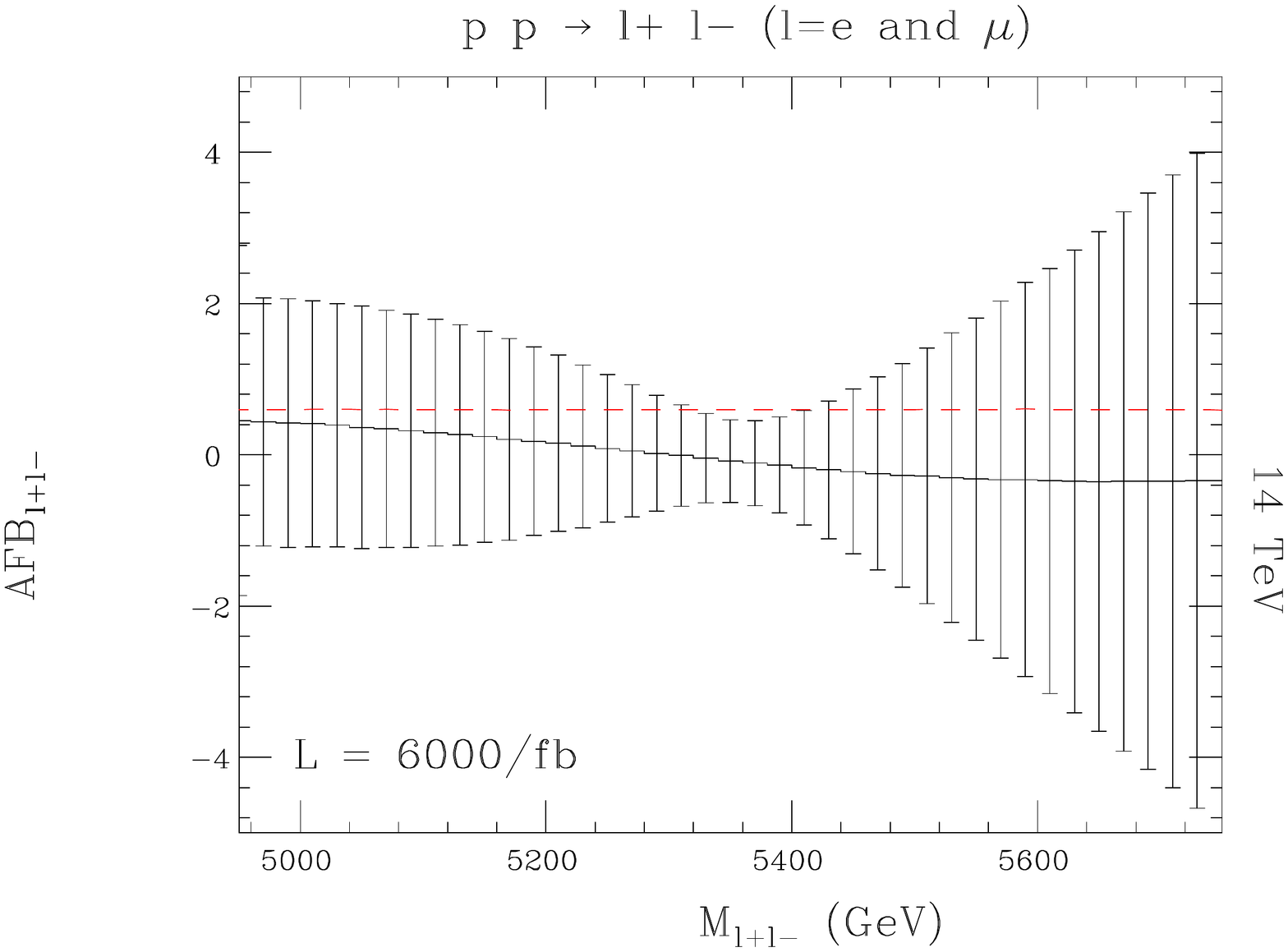} 
\vskip-1cm
\caption{\label{fig:Zp} (Top) Distribution of the cross section vs the di-lepton invariant mass for DY production in the E$_6$SSM (solid black) and SM (dashed red) at the LHC with $\sqrt s=14$ TeV. (Bottom) Distribution of the Forward-Backward Asymmetry vs the di-lepton invariant mass for DY production in the E$_6$SSM (solid black) and SM (dashed red) at the LHC with $\sqrt s=14$ TeV (here, we also give the statistical error for the E$_6$SSM case assuming 6000 fb$^{-1}$ of luminosity). The lightest(heaviest) of the two $Z'$ BPs considered here is on the left(right).} 
\end{figure*}

%%%%%%%%%%

We define a set of Benchmark Points (BPs)  for the study of different signatures. The current LHC limits for a $Z^{\prime}$ in the E$_{6}$SSM are around $4$~TeV \cite{Frank:2020pui} depending on the level of kinetic mixing and whether the decays to Superpartners are allowed or not. Hence,  we  take masses slightly more than $4$~TeV  (BP1, BP2) although we also define a case with $m(Z^{\prime})=5.35$~TeV (BP3) to illustrate the ultimate reach of the High Luminosity LHC (HL-LHC) \cite{Gianotti:2002xx}.

Regarding the inert chargino, we show that the case with $m(\tilde{\chi}_{I}^\pm)=242$~GeV (BP1) could give rise to several signatures, while $m(\tilde{\chi}_{I}^\pm)=273$~GeV (BP2) will be more of a borderline case. The spectra for all  BPs are shown in Tab. \ref{tb:benchmarks}.

%%%%%%%%%%%%%%%%%%%%%%%%%%%%%%%%%%%%%%
\subsection{${\boldmath Z}'$ production}

Concerning a $Z'$ discovery, which is the first E$_6$SSM signature that we  study, 
we demonstrate the sensitivity of the DY channel $pp(q\bar q)\to \gamma,Z,Z'\to l^+l^-$ (with both electrons and muons in the final state, {\it i.e.}, $l=e$ and $\mu$) at the HL-LHC to the aforementioned  BP1 and BP3. In our analysis we set the kinetic mixing small so that the $Z^{\prime}$ can be seen as a narrow resonance.
Specifically, we will look at two extreme configurations, those with lowest(highest) $Z'$ mass, and show that,
despite the rather heavy masses and large widths of the $Z'$s involved in general, {\it i.e.}, approximately 4.212(5.359) TeV and 86(118) GeV,  respectively, 
they afford us with interesting phenomenology. We perform our analysis in respect to both discovery and  characterisation of the $Z'$, by studying simultaneously the $Z'$  cross section ($\sigma$) and   Forward-Backward Asymmetry ($A_{\rm FB})$. The rationale for exploiting both these observables was spelt out in~\cite{Basso:2012ux}--\cite{Accomando:2019ahs}.

We present our results in this respect in Fig.~\ref{fig:Zp}, where we display both $\sigma$ and $A_{\rm FB}$ mapped against the invariant mass of the di-lepton pair, $M_{l^+l^-}$.
In our Monte Carlo (MC) simulation, we adopt the following cuts in lepton transverse momentum and rapidity, respectively:
$p_T^l>20$ GeV and $|\eta^l|<2.5$. Further, given the values of $M_{Z'}$ and $\Gamma_{Z'}$, we limit ourselves to collect results in the 
following di-lepton mass ranges: 3.75(4.95) TeV $< M_{l^+l^-}<$ 4.75(5.75) TeV for the light(heavy) $Z'$. At $\sqrt s=14$ TeV, the SM cross section after such a constraints are enforced is
$5.0\times10^{-3}(3.4\times10^{-4})$ fb while the E$_6$SSM one (including the contribution of the $\gamma,Z$ current alongside the $Z'$ one, together with the relative interference) is $6.1\times10^{-2}(5.2\times10^{-3})$ fb. For the expected HL-LHC luminosity, 6000 fb$^{-1}$ (crucially, combined across ATLAS and CMS\footnote{In fact, we notice that it has become rather customary for the two LHC collaborations to combine their results even in published papers: see, {\it e.g.},  \cite{ATLAS:2017gkv,CMS:2020ezf} for the case of top-quark physics.}), we therefore notice that the SM background is extremely small for the light $Z'$ BP and essentially zero for the heavy one, which means any $Z'$ measurement is essentially background free. However, the  rates for the signal, which is then defined as due to the sum of the pure $Z'$ term squared plus its (extremely small)  interference with the SM one, are not very large either, as they amount to some 330(30) events. Nonetheless, this should allow for the extraction of both the $Z'$ mass value, from the $\sigma$ distribution, and to perform a fit to its coupling parameters, also using the $A_{\rm FB}$ spectrum, which  reveals a rather distinctive shape with respect to the SM one and could be resolved near the $Z'$ peak above and beyond the statistical error. Indeed, we assume that, if a resonance is seen in the former, further run time will be sought, so as to enable a better shaping of the latter. These conclusions are certainly applicable to the light $Z'$ case and possibly to the heavy $Z'$ one too. Altogether then, prospects for profiling a would-be $Z'$ signal are optimistic. On the one hand,  the SM contamination of the cross section spectrum is essentially negligible at the $Z'$ peak (the red dashed curve in the top frames of Fig.~\ref{fig:Zp} has no phenomenological relevance). On the other hand, the diagnostic power of the asymmetry distribution is clearly revealed against the SM contribution in the same mass range (the red dashed curve in the bottom frames of Fig.~\ref{fig:Zp} 
is flat throughout). 
(The code of \cite{Accomando:2019ahs} was used for this part of the analysis.) 

We therefore conclude that DY measurements at the HL-LHC would be sensitive to our BPs, thus contributing to profiling the E$_6$SSM in its extended gauge sector.

%%%%%%%%%%%%%%%%%%%%%%%%%%%%%%%%%%%%%%%%%%%%%%%%%%%%%%%%%%%%%%%%%%%%%%%%%%%%%%%%%%
\subsection{Disappearing tracks of inert charginos}
Next, we study the signature of LL inert charginos based on disappearing tracks at the LHC. When the mass splitting between inert chargino and inert neutralino is of the of order $\simeq 140-350$ MeV, as it is the case in our model (c.f. Tab. \ref{tb:benchmarks}), the chargino becomes a LLP with a lifetime of the order of $\mathcal{O}(0.02\,\textrm{ns})$ and upwards. In such an instance, chargino does not decay promptly but flies through the multiple layers of the tracker, leaving hits that are reconstructed as a disappearing track, which then disappears into the $ \tilde{\chi}_I^0$ MET.

\begin{figure}[t]
\includegraphics[scale=0.25,valign=c]{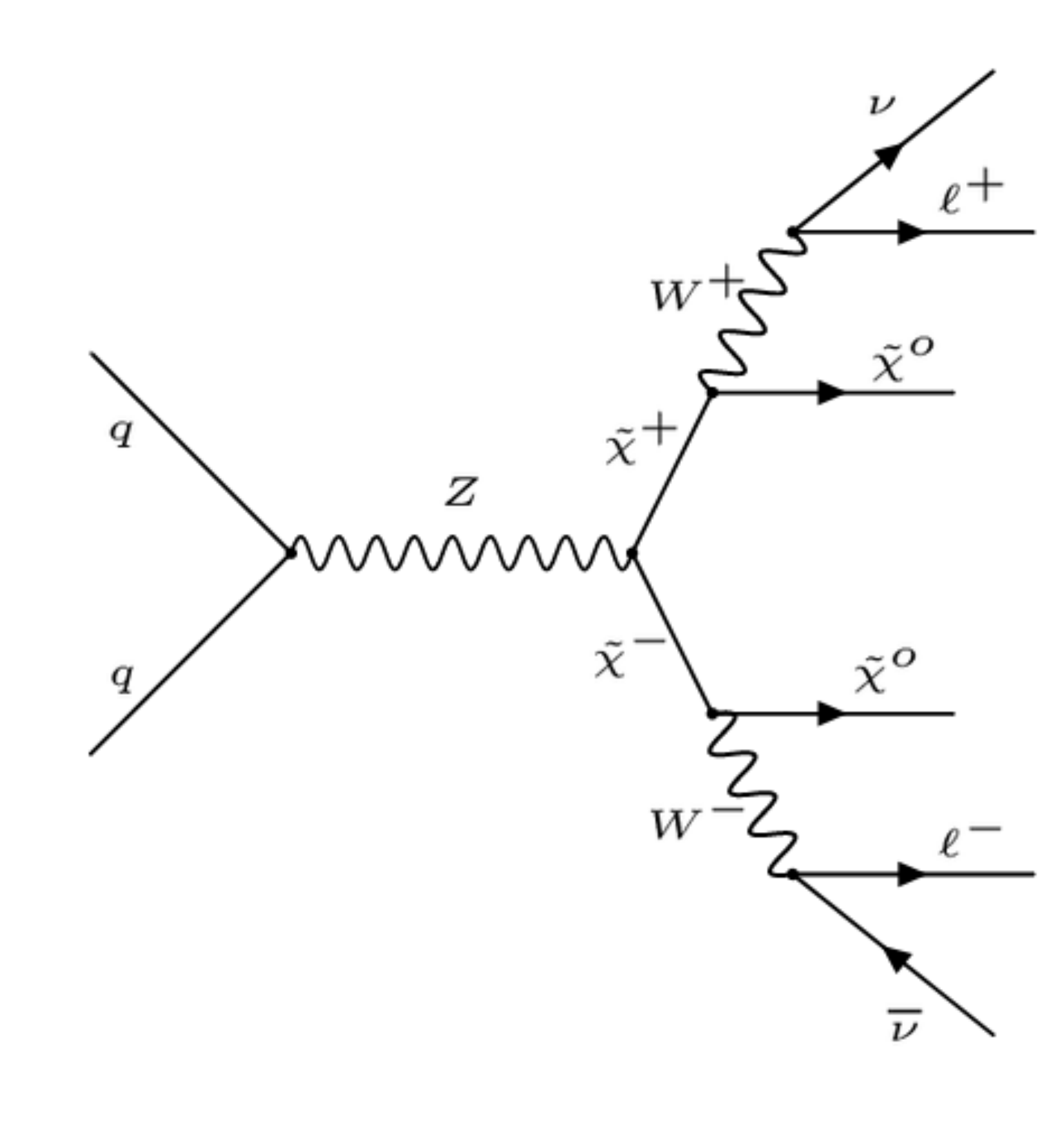}
\hspace{0.6cm}
\includegraphics[scale=0.28,,valign=c]{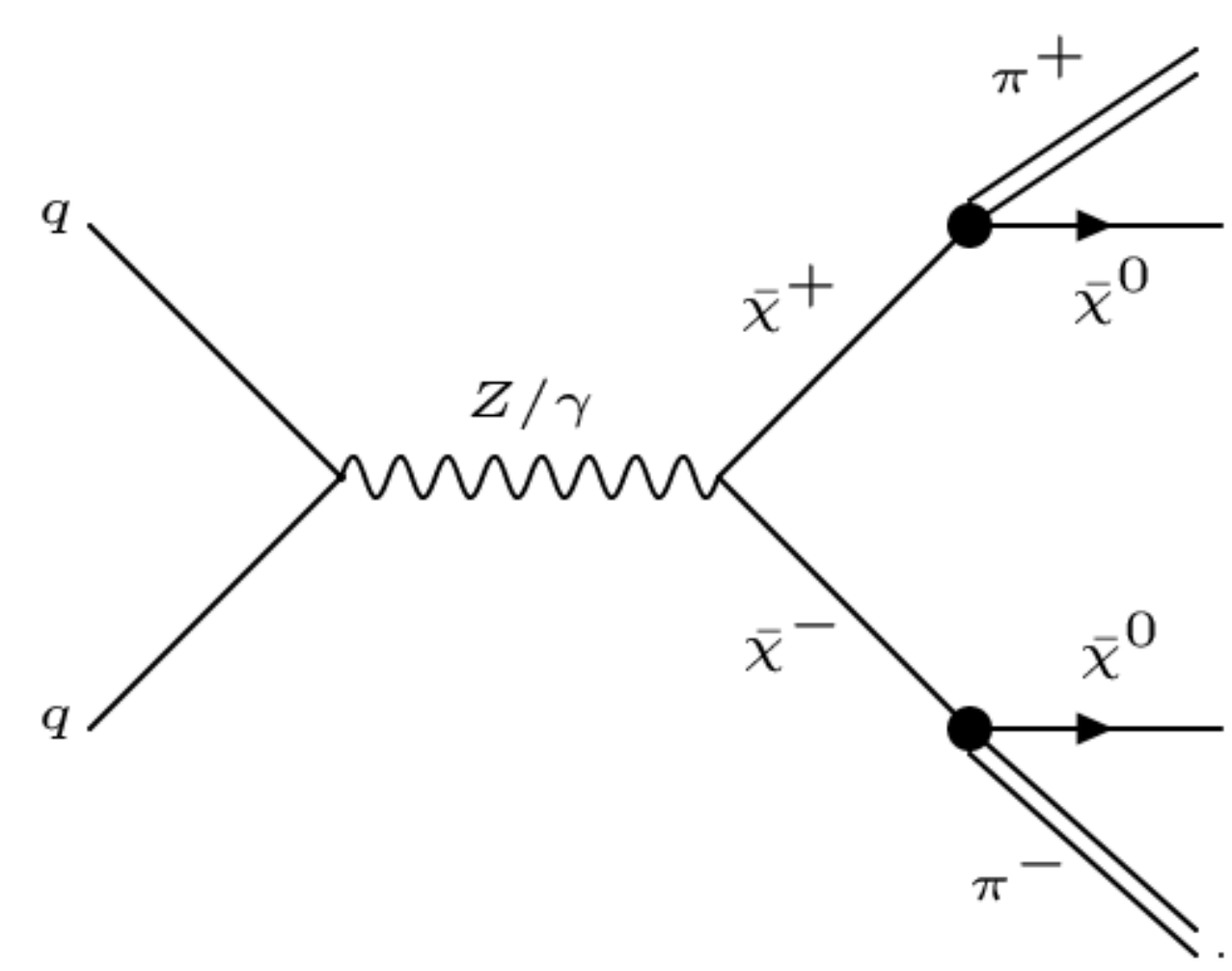}
\caption{Left: Feynman diagram of inert chargino's pair production followed by the decay into inert higgsino DM, soft leptons and Missing Transverse Energy (MET) that generate a disappearing tracks signature. Right: Inert chargino pair production followed by the decay into a pion.} 
\label{Feynman}
\end{figure}

In our study inert charginos can be either pair-produced in the process $p p \to Z/\gamma\to \tilde{\chi}_I^+ \tilde{\chi}_I^-$, or single-produced in association with an inert neutralino, $p p \to W^\pm \to \tilde{\chi}_I^\pm \tilde{\chi}_I^0$. In both cases, charginos  would subsequently undergo a 3-body decay $\tilde{\chi}_I^\pm  \to W^{*\pm}\tilde{\chi}_I^0\to \to  l^\pm \nu \tilde{\chi}_I^0$, with  very soft leptons in the final state. An example of the corresponding Feynman diagram for the inert chargino  pair-production is shown in the left panel of Fig.~\ref{Feynman}.  

\begin{figure}[t]
\includegraphics[scale=0.3]{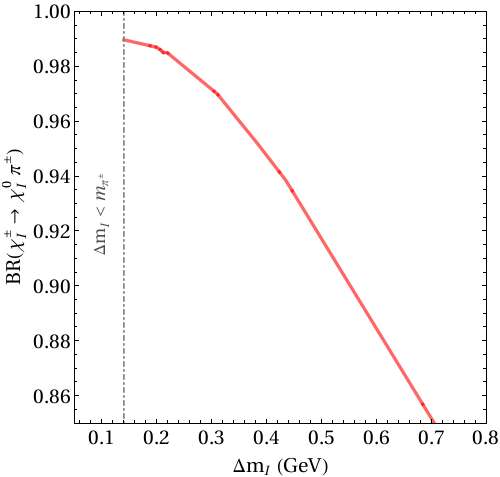}
\caption{$\textrm{BR}(\tilde{\chi}_I^\pm\to \tilde{\chi}_I^0\,\pi^\pm)$ as a function of the inert mass splitting $\Delta m_I$. Below the charged pion mass threshold the only allowed decay channel is a 3-body decay $\tilde{\chi}_I^\pm  \to l^\pm\,\nu\,\tilde{\chi}_I^0$.} 
\label{pionBR}
\end{figure}

It is important to note, however, that when the mass splitting between the LL inert chargino and neutralino, $\Delta m_I$, exceeds but is close to the mass of the charged pion, an additional effect needs to be taken into account. In this regime another decay channel for $\tilde{\chi}_I^\pm$ becomes available, namely $\tilde{\chi}_I^\pm$ decaying to pions via the non-perturbative $W^\pm -\pi^\pm$ mixing term \cite{Thomas:1998wy,Belyaev:2016lok,Fukuda:2017jmk,Belyaev:2020wok} (see the right panel of Fig.~\ref{Feynman}). As a matter of fact, for the mass splitting $\Delta m_I\lesssim 350$ MeV, the decay $\tilde{\chi}_I^\pm\to\tilde{\chi}_I^0\,\pi^\pm$ becomes the dominant decay channel for the inert chargino, as illustrated in Fig.~\ref{pionBR}. The lifetime of $\tilde{\chi}_I^\pm$ reads in this case \cite{Fukuda:2017jmk}
\begin{equation}
\tau_{\tilde{\chi}_I^\pm}\simeq 0.023\,\textrm{ns}\left[\left(\frac{\Delta m_I}{340\,\textrm{MeV}}\right)^3\sqrt{1-\frac{m^2_{\pi^\pm}}{\Delta m_I^2}}\right]^{-1},
\end{equation}
which is usually an order of magnitude shorter than the corresponding lifetime from the 3-body decay \cite{Belyaev:2016lok}.  Note that both pions and leptons are very soft and are typically stopped in the detector.

The ATLAS collaboration performed a series of dedicated disappearing track searches for LL charginos in $pp$ collisions at $\sqrt{s}=8$ and $\sqrt{s}=13$ TeV \cite{ATLAS:2013ikw,Aaboud:2017mpt,ATLAS:2021ttq}, the most recent one with 
$136$ fb$^{-1}$ of data \cite{ATLAS:2021ttq}. The final state is required to present a disappearing track and at least one Initial State Radiation  (ISR)  jet with high $p_T$ to ensure significant amount of MET. The disappearing track search reaches maximum sensitivity  for  charginos with lifetimes of  $\mathcal{O}$(ns). For example, wino-like charginos are excluded at 95\% Confidence Level (CL)  up to 660 GeV, while pure higgsinos  up to 210 GeV. Model independent limits are provided in  \cite{ATLAS:2021ttq} in terms of the 95\% CL upper bound on the visible cross-section $\sigma_{\textrm{vis}^{95\%}}$. The latter can be translated into the upper bound on the observed number of Beyond the SM (BSM) events, $N_{S_{95\%}}=\sigma_{\textrm{vis}^{95\%}}\cdot L$, which at $L=136$ fb$^{-1}$ reads $N_{S_{95\%}}=5$.

\begin{figure}[t]
\includegraphics[scale=0.4]{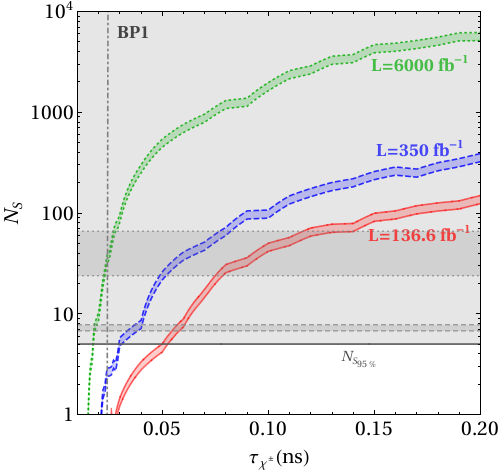}\\
\vspace{0.5cm}
\includegraphics[scale=0.4]{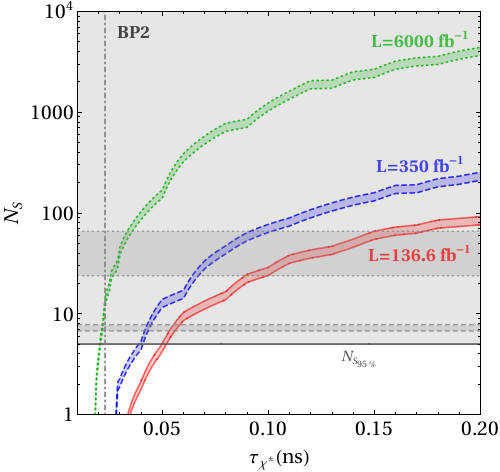}
\caption{Number of expected signal events $N_S$ as a function of the LL $\tilde{\chi}_I^\pm$ lifetime $\tau_{\tilde{\chi}_I^\pm}$ for the inert chargino mass 242~GeV (BP1, upper plot) and 273~GeV (BP2, lower plot), and for three different values of integrated luminosity at the LHC: 136 fb$^{-1}$ (red solid), 350 fb$^{-1}$ (blue dashed), and 6000 fb$^{-1}$ (green dotted).  The vertical dot-dashed line denotes the actual lifetime of the inert chargino. Horizontal solid line corresponds to the 95\% CL upper bounds on the observed number of signal events for the luminosity of 136 fb$^{-1}$. The projected $N_{S_{95\%}}$ ranges are shown as gray-shaded regions bounded by dashed and dotted horizontal lines for the luminosities of 350 fb$^{-1}$ and 6000 fb$^{-1}$, respectively.}
\label{fig:Ns}
\end{figure}

To derive the exclusion limits on the inert LL chargino in our model, we employ the numerical recast tool introduced in \cite{Belyaev:2020wok} and included in the LLP Recasting Repository \cite{LLPRecRep}. The tool takes as a input a  \texttt{.root} file containing the events processed with the detector simulator \textsc{Delphes 3} \cite{deFavereau:2013fsa}. In order to generate the required input, we proceed according to the following numerical receipt. First, we implement the E$_6$MSSM model in \texttt{Sarah v4.14} \cite{Staub:2008uz}--\cite{Staub:2013tta} and pass the corresponding {\tt UFO} files to \textsc{MadGraph5\_aMC@NLO} \cite{Alwall:2014hca}. \textsc{Pythia 8} \cite{Sjostrand:2007gs} is then used for showering, and the hadronisation products are passed to \textsc{Delphes3}. The disappearing tracks  recast tool gives as an output the number of signal events predicted by the model as a function of the varying  LLP lifetime, with an overall error of this determination of around 20\%. This number of signal events should then be compared to the $N_{S_{95\%}}=5$ determined by the ATLAS analysis  \cite{ATLAS:2021ttq} to decide whether the model point under study can be excluded at 95\% CL or not.

We are also interested in providing a potential future reach of the disappearing track searches for higher luminosities.
To this end, let us recall that the counting-experiment likelihood is given by the Poisson distribution convolved 
with an additional function that takes into account the uncertainty in the background determination,      
\begin{equation}\label{eq:like}
\mathcal{L}=\frac{1}{\sqrt{2\pi}\,\delta B}\int dB'\,\frac{e^{-(S+B')}(S+B')^{O}}{O!}\, e^{-\frac{(B'-B)^2}{2\,\delta B^2}}\,.
\end{equation}
In the above, $S$ indicates the signal yield, $O$ is the observed number of events, $B$ the expected SM background, and $\delta B$ the experimental estimate of the systematic uncertainty. Defining the test statistics $\Delta\chi^2$ as $\Delta\chi^2=-2\log(\mathcal{L}/\mathcal{L}_{0})$, where $\mathcal{L}_{0}$ corresponds to the background-only hypothesis, one can obtain the 95\%~CL limit on the number of signal events by requiring $\Delta\chi^2=3.84$. 

Eq.~\ref{eq:like} can now be used to derive projections for the future reach of the disappearing track searches by setting $O=B$. Under the assumption that the experimental analysis \cite{ATLAS:2021ttq} at higher luminosities is not modified in terms of the event reconstruction and selection strategy, the background yield would scale like $L/\textrm{136}\,\textrm{fb}^{-1}$. 
On the other hand, the corresponding $N_{S_{95\%}}$, derived from Eq.~\ref{eq:like}, will strongly depend on the assumption about the future systematic uncertainty of the background yield. To account for the possibility that the future background determination by the experimental collaboration will be more precise, we are going to consider two limiting cases: a) systematic uncertainty $\delta B$ of the same order as in  \cite{ATLAS:2021ttq},  and b) systematic uncertainty reduced to around $\delta B=1\%$ of the total background yield. As a result, the projected values of signal events at 95\% CL read $N_{S_{95\%}}\in [6.8-7.8]$ for the luminosity of $350$ fb$^{-1}$, and $N_{S_{95\%}}\in [24-66]$ for 6000 fb$^{-1}$. 

In Fig.~\ref{fig:Ns}  we show the number of signal events, $N_S$, for the LL inert charginos with masses 242 GeV (BP1, upper panel) and 273~GeV (BP2, lower panel), assuming different chargino lifetimes. Various colours and styles correspond to integrated luminosities of 136 fb$^{-1}$ (red solid), 350 fb$^{-1}$ (blue dashed) and 6000 fb$^{-1}$ (green dotted). The width of the bands indicate the 20\% error in the signal determination. The 95\% CL exclusion bound on the number of signal events by ATLAS \cite{ATLAS:2021ttq} is shown as a gray horizontal line. The parameter space above this line is excluded. 
The corresponding $N_{S_{95\%}}$ ranges for the luminosities of 350 fb$^{-1}$ and 6000 fb$^{-1}$ are shown as gray-shaded regions bounded by dashed and dotted horizontal lines, respectively. The vertical dot-dashed line corresponds to the actual lifetime of the inert chargino in our model, $\tau_{\tilde{\chi}_I^\pm}=0.024 (0.023)$~ns for BP1 (BP2). 

One can observe that, while the inert chargino predicted by our model is not yet tested in the disappearing tracks searches, there is a chance of finding statistical hints about its existence at the HL-LHC with the expected (combined) luminosity of 6000 fb$^{-1}$. For example, BP1 can be excluded if the experimental uncertainties in the background determination are reduced by a factor of 2 w.r.t the present analysis \cite{ATLAS:2021ttq}, {\it i.e.}, to $\sim 12\%$. BP2, on the other hand, is practically the heaviest inert chargino we can be able to exclude, which would also require some further improvements on the experimental side in the efficiency of the background rejection. 

Therefore, the disappearing tracks signature could be another promising way of testing (and hopefully discovering) the E$_6$SSM model at the CERN machine.

%%%%%%
\subsection{In-flight conversion}

It is also possible to generate the inert charginos via the $Z^{\prime}$ portal. In such a case the charginos have a large momentum $p(\tilde{\chi}_{I}^{\pm})\geq m(Z^{\prime})/2$, so for our benchmarks $\gamma\simeq m(Z^{\prime})/2m(\tilde{\chi}_{I}^{\pm})> 7$. Hence $\beta\gamma$ is clearly larger than than $3$, so the ionisation losses of inert charginos correspond to a minimally ionising particle. Thus the searches of LLPs based on a different ionisation rate (\textit{e.g.} \cite{CMS:2016kce}) are not sensitive to this signal. In contrast,  the boost can be so large that the leptons may be reconstructed in the electromagnetic calorimeters and muon detectors. In practice this looks as if a chargino had been converted to an electron or a muon in flight as the momentum of the charged lepton is almost aligned to that of the chargino. If the inert chargino decays to a neutralino and hadrons,  the latter are so soft that they will never be reconstructed as a jet.

The signature is a soft electron or muon with a curved track that meets a reasonably long ($\sim 5$~cm) nearly straight track. We select muons with $p_{T}>3$~GeV and electrons with $p_{T}>5$ ~GeV and require them to be separated from other objects within $\Delta R =0.3$. For the leptons of BP1 (BP2) we have an average transverse momentum of $4.0$~GeV ($3.9$~GeV) and an average chargino decay length of $59$~mm ($43$~mm). We used a dedicated version of Delphes and MadAnalysis to treat the displaced vertices \cite{DVpackage}. If the chargino does not reach the innermost layer of the pixel detector, the curved lepton track will just simply not point towards any primary vertex.

The challenge in discovering such a signature lies in triggering it. The soft ($p_{T}^l\sim 5$~GeV) electron or muon will not be sufficient for triggering, so the best chances are by using a trigger based on MET. We show our results based on the assumption that events with $\slashed{E}_{T}>150$~GeV can be triggered with a nearly $100\%$ efficiency \cite{CMS:2020cmk}. If triggering could be based on tracks only, obviously the chances would be better. Such an option is not available currently, but there are plans to include triggering based on tracker data to the HL-LHC trigger. The figures in Table \ref{tb:dispvertex} are based on perfect triggering and should be multiplied with the triggering efficiency (which obviously currently is unknown).

The pixel detectors of CMS and ATLAS have four layers each \cite{CMSTrackerGroup:2020edz}--\cite{ATLAS:2010ojh}. For the benchmarks BP1 and BP2 we present in Tab. \ref{tb:dispvertex} the number of inert charginos that lead to a displaced lepton reaching each of the layers of the pixel detector using either the assumption that displaced vertices could be triggered based on tracks only or using the $\slashed{E}_{T}$ trigger.

\begin{table}
\begin{tabular}{c|c|c|c|c}
\hline
\hline
Layer & BP1 & BP1, MET & BP2 & BP2, MET\\
\hline
 & $11.6$ & $3.0$ & $7.2$ & $1.8$\\
1 & $6.7$ & $1.4$ & $3.3$ & $1.0$\\
2 & $3.7$ & $1.0$ & $1.5$ & $0.6$\\
3 & $1.7$ & $0.5$ & $0.6$ & $0.2$\\
4 & $0.7$ & $0.15$ & $0.11$ & $0.05$\\
\hline
\hline
\end{tabular}
\caption{The number of charginos leading to displaced leptons, which reach each of the layers of the pixel detector of CMS or ATLAS. The values are given for a total integrated luminosity of $6000$~fb$^{-1}$. In the columns with MET a requirement of $\slashed{E}_{T}>150$~GeV has been imposed.\label{tb:dispvertex}}
\end{table}

We may see that even at the HL-LHC the event rates would be very small, but the SM backgrounds for such events are practically zero, especially if we have more than one hit in the pixel detector. Such a signature could be more promising at colliders with higher energies as the production of heavier $Z^{\prime}$ bosons could lead to more charginos having a high enough boost to produce leptons that reach the calorimeters and muon detectors. For instance if the high-energy upgrade of the LHC would operate with $\sqrt{s}=27$~TeV, the production cross section of $Z^{\prime}$ bosons would increase by a factor of $50$, which would allow to probe larger $Z^{\prime}$ and chargino masses. At higher chargino masses the $W/Z$ mediated processes become more off-shell, so the relative importance of this channel might increase.
%%%%%%%

\subsection{DM signatures}

We now discuss also the possibilities of DM direct detection experiments. As shown in Tab. \ref{tb:DD-LLPtest2},  our BPs survive the constraints of current experiments \cite{Aprile:2018dbl,Aprile:2019dbj} but, since future ones are expected to improve the bounds by more than an order of magnitude \cite{DARWIN:2016hyl}, the active component can be discovered. The inert component has a tiny DD cross section that is below the neutrino background. We calculated the spin-independent (SI) and spin-dependent (SD) DM scattering cross section, $\sigma^{\rm SI}_{\rm proton}$ and $\sigma^{\rm SD}_{\rm proton}$ respectively, using \textsc{micrOMEGAs v5.2.4} \cite{Belanger:2008sj, Belanger:2014vza}.

The interactions between DM and the quark sector is mainly mediated by the $Z'$ exchange, where the effective interaction is given by
\be%
{\cal L_{\text{eff}}} = f_q \overline{\tilde{\chi}} \tilde{\chi} \, \bar{q} q+b\alpha_{s}\overline{\tilde{\chi}} \tilde{\chi}G^{\mu\nu a}G^{a}_{\mu\nu}, %
\label{scalar}
\ee%
where $q$ stands either for proton or neutron, $f_q \propto g^2_{N}/M_{Z'}^2$ comes from the $Z^{\prime}$ mediated process and $b$ is Higgs-gluon coupling induced by the heavy quark loops. 
%The effective coupling of $\tilde{\chi}$  with protons and neutrons $f_p$, $f_n$ can be computed in terms of $f_u$, $f_d$ and $b$.
The zero momentum transfer scalar cross section of the higgsino scattering with the nucleus is given by \cite{Belanger:2008sj}:
\be%
\sigma^{\rm SI}_0 = \frac{4 m_r^2}{\pi} \left(Z f_p + (A-Z) f_n \right)^2,
\ee%
where $Z$ and $A-Z$ are the number of protons and neutrons, respectively, $m_r=m_N m_{\tilde{\chi}_1}/(m_N+m_{\tilde{\chi}_1})$, and $m_N$ is the nucleus mass. Thus, the differential scalar cross section for non-zero momentum transfer $q$ can be written as
\be%
\frac{d \sigma_{\rm SI}}{dq^2} = \frac{\sigma^{\rm SI}_0}{4 m_r^2 v^2}F^2(q^2),\; 0< q^2 < 4 m^2_r v^2,
\ee%
where $v$ is the velocity of the lightest neutralino and $F(q^2)$ is the relevant Form Factor (FF).
Therefore, the SI (scattering) cross section of the LSP with a proton is given by
\be
\sigma_{\rm SI}^p=\int_0^{4 m^2_r v^2}\frac{d \sigma_{\rm SI}}{dq^2}\big{|}_{f_n=f_p} dq^2.
\ee

%In figure \ref{SI-SD}, we display the SI scattering cross section of the active and inert higgsino LSP with a proton after imposing the relic abundance constraints. 

The SD interaction of a DM candidate stems solely from the quark axial current: 
$$ a_N  \bar{\chi} \gamma^\mu \gamma_5 \chi ~ \bar{N} \gamma_\mu \gamma_5 N,$$ 
where $a_N = \sum_{q=u,d,s} d_q \Delta_q^{(N)}$, with $d_q$  the effective quark level axial-vector
and pseudoscalar couplings and $\Delta_q^{(N)}$ is given via 
$\Delta_u^{(p)}=\Delta_d^{(n)}= 0.77 $, $\Delta_d^{(p)}=\Delta_u^{(n)}= -0.40 $, and $\Delta_s^{(p)}=\Delta_s^{(n)}= -0.12 $ \cite{Belanger:2008sj}.   
In this case, the SD (scattering) cross section of DM-nucleus is given by 
\be 
\sigma_{\rm SD} = \frac{16}{\pi} m_r^2 a_N^2 J_N (J_N +1), 
\ee
where $J_N$ is the angular momentum of the target nucleus. In case of the proton target, $J_N =1/2$.

\begin{table}[!t]
\begin{tabular}{l l c c| c c}
\hline
\hline
& DMC  & $\sigma^{\rm SI}_{\rm proton}$ & $\sigma^{\rm SI}_{\rm proton}$ & XE-1T SI & XE-1T SD\\
\hline
BP1 & active & $8.38\times 10^{-10}$ & $7.57\times 10^{-7}$ & $9.6\times 10^{-10}$ & $4.3\times10^{-3}$\\
BP1 & inert & $1.18\times 10^{-14}$ & $4.71\times 10^{-11}$ & $2.0\times 10^{-10}$ & $9.5\times 10^{-4}$\\
\hline
BP2 & active & $9.28\times 10^{-10}$ & $8.15\times 10^{-7}$ & $9.4\times 10^{-10}$ & $4.3\times10^{-3}$\\
BP2 & inert & $1.51\times 10^{-14}$ & $3.85\times 10^{-11}$ & $2.3\times 10^{-10}$ & $1.1\times 10^{-3}$\\
\hline
\hline
\end{tabular}
\caption{Direct detection cross sections (in pb) for the two DM Candidates (DMCs) and the corresponding limits from XENON-1T for the spin-independent  \cite{Aprile:2018dbl} and  spin-dependent  \cite{Aprile:2019dbj} case.\label{tb:DD-LLPtest2}}
\end{table}

\begin{figure}[h!]
 \includegraphics[scale=0.65]{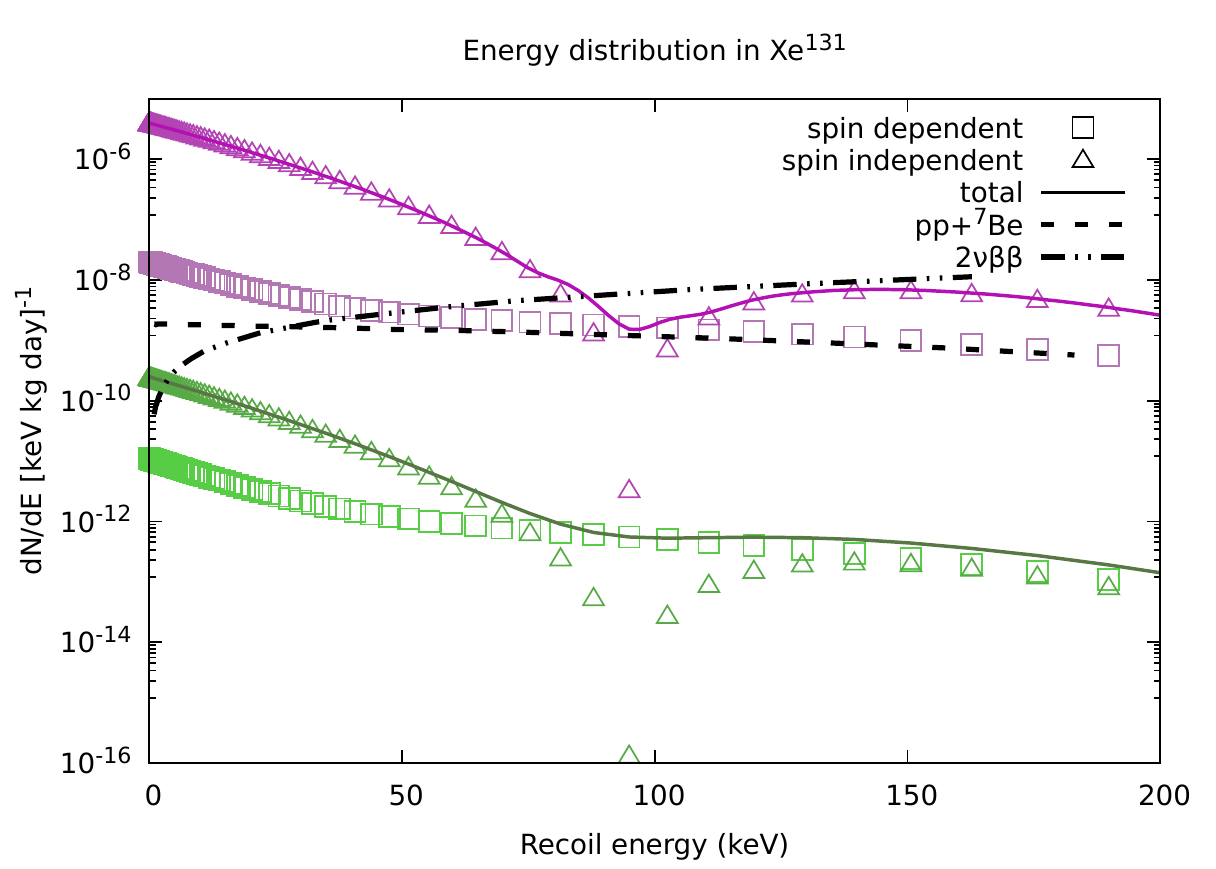}
 \includegraphics[scale=0.65]{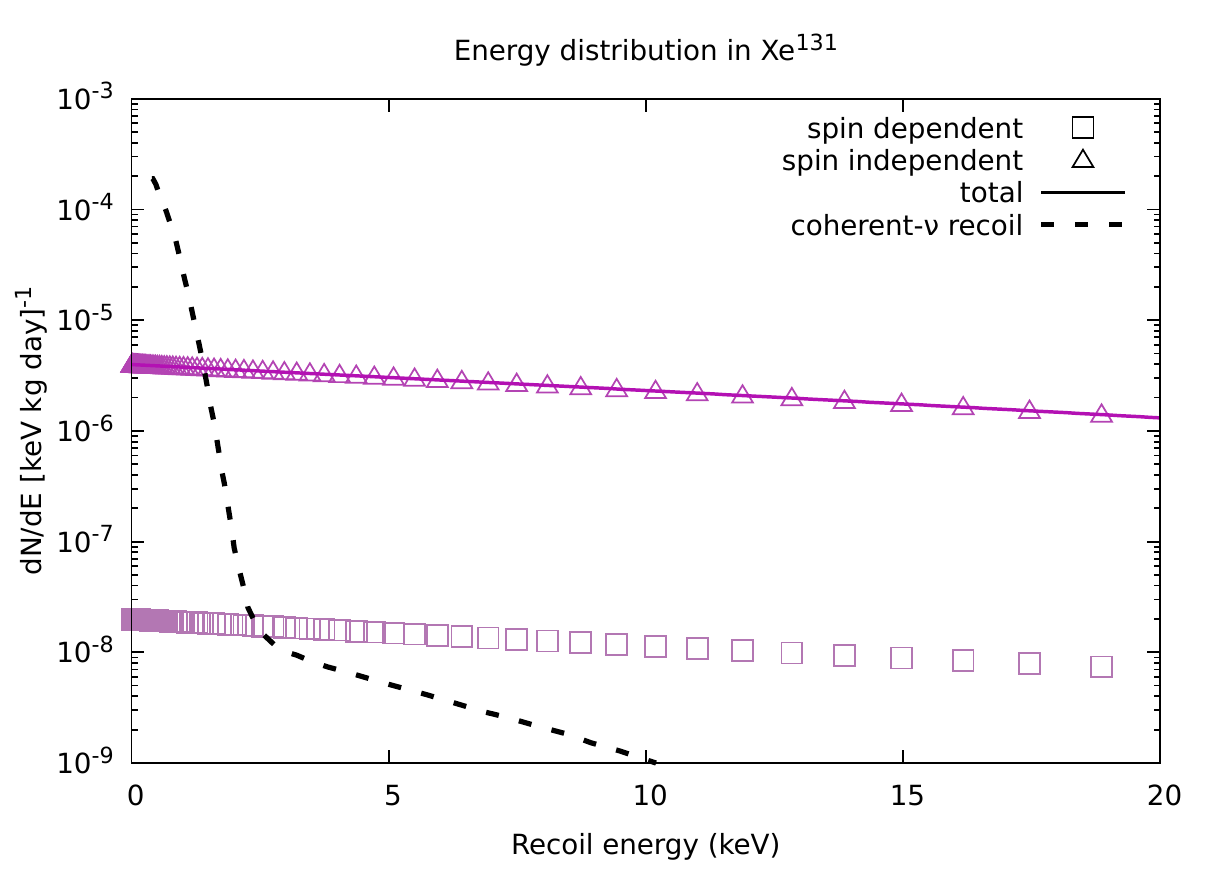}
 \includegraphics[scale=0.65]{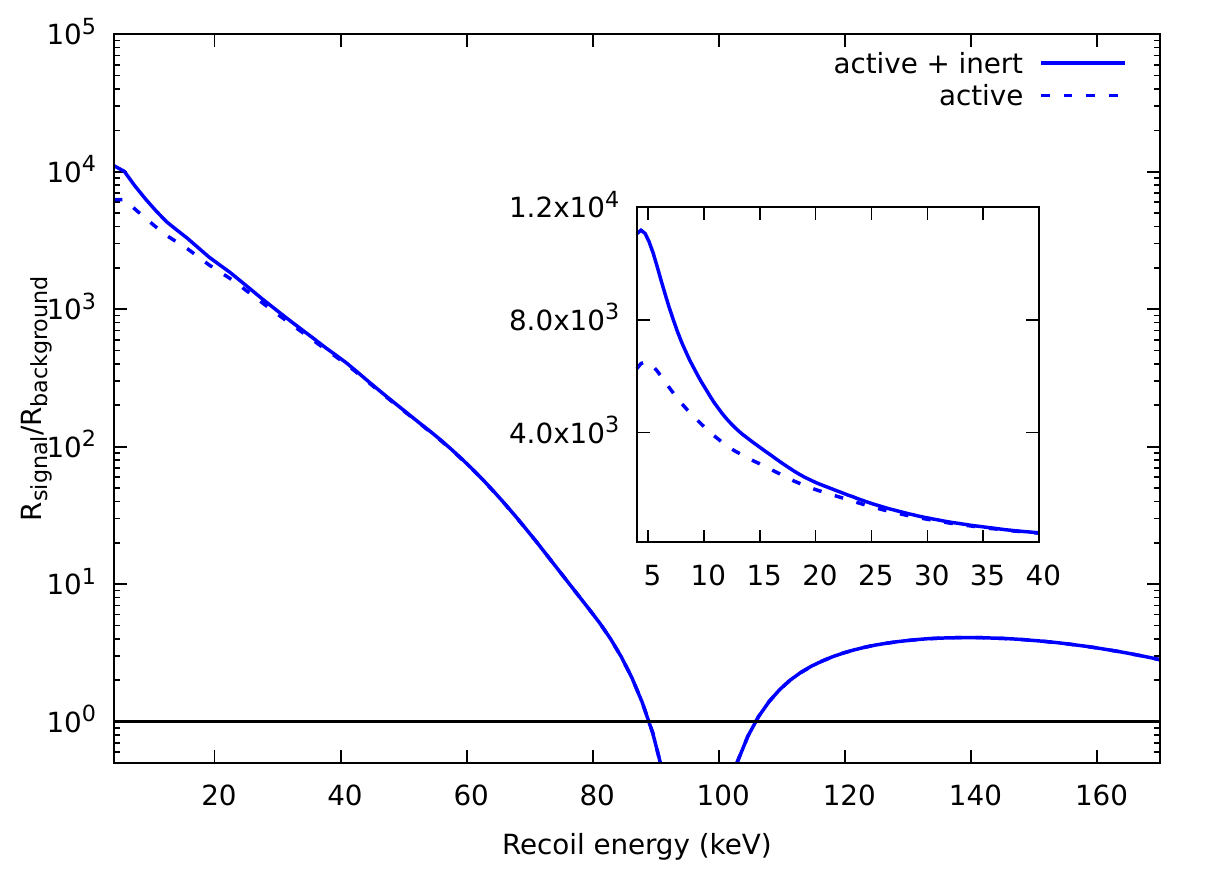}
	\caption{\small (Top) Nuclear recoil spectrum for the selected BP2 in a liquid Xe detector (BP1 results are the same), for both the active (purple) and inert (green) neutralino. Also shown (black) are the expected backgrounds \cite{DARWIN:2016hyl,Baudis:2013qla}. (Middle) We show the coherent neutrino nuclear recoil scattering, which affects in the low recoil energy spectrum. (Bottom) Ratio of active plus inert neutralino (solid) and active neutralino only (dotted) signal rates to the total background ones as obtained from the top plot. In addition, a detailed view of the difference in linear scale is also displayed.}
		\label{Xebkg}
\end{figure}

At the top of Fig. \ref{Xebkg}, we show the nuclear recoil spectrum for one benchmark, the results for the others are practically the same. We can see that the active component is visible, while the inert signal would be lost as it lies below the so-called \emph{neutrino floor}. Neutrino interactions in DM direct detection experiments constitute a significant background that imposes a lower bound on the sensitivity of these experiments\cite{DARWIN:2016hyl,Baudis:2013qla,Gaspert:2021gyj}. These neutrino fluxes that can arrive at the detector are most of solar origin, mainly coming from $pp$-reaction (proton fusion reaction at the centre of the Sun), and the $^7$Be-neutrinos (from the electron capture reaction $^7$Be $+ e^- \to ^7$Li $+ \nu_e$) \cite{Gaspert:2021gyj}. 
Other elastic neutrino-electron interactions which constitute a background that can potentially compete with a DM signal come from the gas impurities of the liquid xenon itself, such as $^{85}$Kr and $^{222}$Rn, but mainly from double beta decays of $^{136}$Xe. 
To model the shape of these backgrounds, one needs to make many assumptions ranging from the Standard Solar Model to the systematic measurement errors, including the level of discrimination between nuclear and electronic recoils.
In here, we adopt the assumptions of \cite{Baudis:2013qla}.

In the middle plot of Fig. \ref{Xebkg}, we show the nuclear-neutrino coherent scattering, which is the sum of mainly solar neutrinos ($^8$B), and also atmospheric and diffuse supernovae neutrinos \cite{Baudis:2013qla}. 
The SM predicts the cross section of this background, but it has not yet been observed.
The highest event rate comes from solar neutrinos, but it peaks at $~1.2$keV. 
Therefore, it is an essential background for WIMP masses below 10 GeV, which signals mimic the $^8$B neutrinos and produce no trace above $~2$ keV.
For larger WIMP masses, the trace is distinguishable. 
As it is a nuclear recoil event, the rates are higher than neutrino-electron recoil shown at the top, and for very low recoil energies, it would be larger than the active DM signal. 

At the bottom of Fig. \ref{Xebkg}, we show the difference of the rate event signal (using the spin-independent nuclear recoil) if in the model there are two DM candidates (active plus inert) or if the active is the only DM candidate. 
As discussed in \cite{Khalil:2021ycm}, the recoil spectrum has some sensitivity to the DM particle mass, so for our benchmarks, the shape of the spectrum would tell us that we found a different DM candidate than the inert one found at the colliders.

\subsection{MET searches}

We finally remark that, if the active neutralinos are heavy, traditional SUSY searches for, say, stops or gluinos are not sensitive as the amount of $\slashed{E}_{T}$ will be too similar to SM backgrounds and hence discovering the colored superpartners will be difficult. In contrast,  if the active neutralino is light (too light to saturate the relic density bound on its own) and the inert one heavy, the disappearing track signature from the inert sector will vanish, but traditional SUSY cascades will have more $\slashed{E}_{T}$ and the searches for standard SUSY cascades may be sensitive.
\\

%%%%%%%%%%%%%%%%%%%%%%%%%%%%%%%%%%%%%%%%%%%%%%%%%%%%%%%%%%%%%%%%
\section{Conclusions}

In conclusion, we have proposed here a set of signatures  of a simplified  E$_6$SSM that may emerge in a variety of space and ground experimental data collected 
by existing  facilities  in the near future as a blueprint of a specific BSM construction based on non-minimal SUSY with a string theory origin. We have done so by using a few BPs as representative of a rather narrow, yet not particularly fine-tuned, region of parameter space. We look forward to dedicated experimental analyses of our model. In fact, further signatures of this simplified  E$_6$SSM may also be accessed at proposed future machines, like an  $e^+e^-$ collider \cite{Khalil:2021ycm}.\\

\section*{Acknowledgments} The work of SK was
partially supported by the STDF project 37272. SM  is  financed  in part through the NExT Institute and the STFC consolidated Grant No.  ST/L000296/1.  HW acknowledges financial  support  from  the  Finnish  Academy  of  Sciences and Letters and STFC Rutherford International Fellowship scheme (funded  through  the  MSCA-COFUND-FP  Grant No. 665593). KK and DRC are supported by the National Science Centre (Poland) under the research Grant No. 2017/26/E/ST2/00470. The  authors  acknowledge  the  use  of  the IRIDIS High Performance Computing Facility and associated support services at the University of Southampton, in the completion of this work. The use of the CIS computer cluster at the National Centre for Nuclear Research in Warsaw is gratefully acknowledged.

\end{document}